# WHAT IS SGR A*?

*The Starved Black Hole in the Center of the Milky Way*[1]

H. FALCKE

*Max-Planck-Institut für Radioastronomie (MPIfR)*
*Auf dem Hügel 69, D-53121 Bonn, Germany*
*E-mail: HFalcke@mpifr-bonn.mpg.de*

## 1. Introduction

Sgr A* is the unique 1 Jy flat spectrum radio point source located at the dynamical center of the Galaxy and in the very center of the central star cluster (Eckart et al. 1993). Due to its unusual appearance it has long been speculated that this source is powered by a supermassive black hole – an object whose presence has been suspected to reside in the nuclei of many other galaxies as well. Its mass is believed to be as large as $M_\bullet \sim 2 \cdot 10^6 M_\odot$ (e.g. Genzel & Townes 1987) while a lower limit of $M_\bullet > 200 - 2000 M_\odot$ can be inferred from the low proper motion of Sgr A* (Backer – this volume).

The enormous increase in observational data obtained for Sgr A* in recent years has enabled us to develop, compare and constrain a variety of models for the emission characteristics of this source. Because of its relative proximity and further observational input to come Sgr A* may therefore become one of the best laboratories for studying supermassive black hole candidates and basic AGN physics. This paper briefly summarizes our current understanding of this enigmatic radio source.

## 2. Observational Input

### 2.1. RADIO-SUBMM SPECTRUM

The radio spectrum of Sgr A* has been extensively studied in the range 1-600 GHz where it mostly shows substantial variability. As there are only very few quasi-simultaneous flux density measurements available (see Wright & Backer 1993) an exact description of the radio spectrum is very uncertain at the moment. An averaged spectrum combined of various data sets available in the literature (Duschl & Lesch 1994) may be fitted by a single powerlaw with spectral index $\alpha \sim 1/3$ ($S_\nu \propto \nu^\alpha$). However, it appears as if the submm regime is less variable than the radio regime (Zylka et al. 1994) and there might even be a weak submm-excess (Zylka et al. 1992; compare also Rogers et al. 1994 with Zylka et al. 1994). Sgr A* is not seen at IR wavelength and hence the spectrum must cut-off towards 12$\mu$m (Zylka et al. 1992; Gezari – this volume).

---

[1] Invited Review – to appear in the proceedings of the IAU Symp. 169 "Unsolved Problems of the Milky Way", L. Blitz (ed.), Kluwer Academic Press

2.2. HIGH ENERGIES

Evidence has grown substantially that Sgr A* is also an X-ray emitter. Art-P/GRANAT detected an x-ray source coinciding with the position of Sgr A* within 40″ (Sunyaev et al. 1991). It shows variability within a factor 2 over a period of several months. The identification as Sgr A* is corroborated by a ROSAT detection of this source with a positional uncertainty of only 10″ (Predehl & Trümper 1994). The ROSAT flux, however, was lower than expected from the Art-P measurements and prompted the interpretation of additional intrinsic absorption in Sgr A*. The spectrum in the Art-P band (4-20 keV) is a hard powerlaw with $\alpha \sim -0.6$ and breaks already in the range 35-100 keV (Goldwurm et al. 1994). There is also a gamma-ray detection of the Galactic Center (GC) with EGRET (Mattox et al. 1992) but at present it is not clear whether this is a point source or extended emission.

2.3. LUMINOSITY CONSTRAINTS

The bolometric optical-UV luminosity of Sgr A* can be estimated from the fact that a luminous point source should contribute to the heating of the surrounding dust and thus be visible in submm-IR data (Falcke et al. 1993a) – which is not the case. Hence, we estimated that Sgr A* can not be very luminous with $L_{\rm UV} \leq$ a few $10^5 L_\odot$. Recently Zylka et al. (1994) have updated their submm measurements of the Sgr A region and concluded from the low temperature gradients in the dust that dust heating can not be dominated by a single point source but is more likely due to a cluster of luminous stars (e.g. Krabbe et al. 1991). A lower limit derived from the claimed detection of Sgr A* at NIR wavelengths (Eckart et al. 1992) has become uncertain as this source was now resolved into a cluster of stars (Genzel – this volume) making it difficult to identify Sgr A* with the present uncertainties between the radio and optical reference frame. Only the model dependent lower limit of $L_{\rm UV} > 10^4 L_\odot$ derived from the radio emission (Falcke et al. 1993b) remains still valid.

2.4. SOURCE SIZE

The mm-submm size of Sgr A* is constrained at least within one order of magnitude. From the absence of refractive scintillation Gwinn et al. (1991) have argued that Sgr A* must be larger than $10^{12}$ cm at $\lambda 1.3$ and $\lambda 0.8$ mm. Krichbaum et al. (1993 & 1994) obtained source sizes for Sgr A* of $4.2 \cdot 10^{13}$ cm at 86 GHz and $9.5 \cdot 10^{13}$ cm at 43 GHz with VLBI – the latter well above the expected scattering size as extrapolated from lower frequencies. This claim is challenged by Rogers et al. (1994) who only get $2 \cdot 10^{13}$ cm at 86 GHz in an experiment with a factor 2 shorter baseline. Krichbaum et al. (1993) also found additional weak components and a somewhat elongated source structure at 43 GHz VLBI not seen by Backer et al. (1993). A possibility to reconcile the results could be source variability and elongation of the internal structure which would lead to different sizes if observed with differently oriented baselines. It will be very interesting to see the results of further mm-VLBI experiments.

## 3. Properties of the radio source

### 3.1. A HOMOGENOUS BLOP?

Recent submm measurements (see Zylka et al. 1994) indicate that the radio spectrum of Sgr A* continues up to several hundred GHz with peak fluxes around 3.5 Jy and a sharp cut-off towards the IR. The submm spectrum can no longer be explained by thermal dust emission as this would require extremely cold dust ($\sim$ 15K) which is very unlikely because of the intense (stellar) radiation field in the Galactic Center. To explain the flat submm spectrum with synchrotron emission one needs either a combination of self-absorbed components (requiring high compactness) or an electron distribution where the bulk of the electron energy is concentrated in a narrow energy interval. The latter could be either a very flat electron powerlaw distribution ($dN/d\gamma \propto \gamma^{-p}$) with $p < 1/3$ and sharp high-energy cut-off, a steep powerlaw with low-energy cut-off, a monoenergetic (e.g. an electron beam) or a thermal distribution.

Duschl & Lesch (1994, also this volume) suggested that the radio emission of Sgr A* can simply be explained with a single homogenous blob of monoenergetic electrons. Although this can not be quite true in its most rigorous formulation, as argued below, one can use this approach to get a fairly good idea of the basic parameters of the Sgr A* radio source: the required model parameters are the magnetic field $B$, the Lorentz factor $\gamma_e$, the electron density $n_e$, the volume $V = \pi R^2 Z$ (assumed to be cylindric) and the distance set to 8.5 kpc. On the other side we have three measurable input parameters: the peak frequency $\nu_{\max} \sim \nu_c/3.5$ of a monoenergetic synchrotron spectrum, the peak flux $S_{\nu_{\max}}$ and the VLBI source size (see above). A fourth parameter can be gained if one assumes that magnetic field and relativistic electrons are in equipartition, i.e. $B^2/8\pi = k n_e \gamma_e m_e c^2$ with $k \sim 1$. With this condition we obtain (averaged over pitch angle) that

$$\gamma_e = 326 \; k^{1/7} \left(\frac{F_{\nu_{\max}}}{3.5\mathrm{Jy}}\right)^{-1/7} \left(\frac{\nu_{\max}}{10^{12}\mathrm{Hz}}\right)^{3/7} \left(\frac{R}{10^{13}\mathrm{cm}}\right)^{2/7} \left(\frac{Z}{4\cdot 10^{13}\mathrm{cm}}\right)^{1/7}$$

$$B = 10 \; \mathrm{G} \; k^{-2/7} \left(\frac{F_{\nu_{\max}}}{3.5\mathrm{Jy}}\right)^{2/7} \left(\frac{\nu_{\max}}{10^{12}\mathrm{Hz}}\right)^{1/7} \left(\frac{R}{10^{13}\mathrm{cm}}\right)^{-4/7} \left(\frac{Z}{4\cdot 10^{13}\mathrm{cm}}\right)^{-2/7}$$

$$n_e = \frac{1.4 \cdot 10^4}{\mathrm{cm}^3} k^{2/7} \left(\frac{F_{\nu_{\max}}}{3.5\mathrm{Jy}}\right)^{5/7} \left(\frac{\nu_{\max}}{10^{12}\mathrm{Hz}}\right)^{-1/7} \left(\frac{R}{10^{13}\mathrm{cm}}\right)^{-10/7} \left(\frac{Z}{4\cdot 10^{13}\mathrm{cm}}\right)^{-5/7}.$$

Apparently the 'non'-equipartition parameter $k$ enters only weakly and as long as one is not very far from equipartition the parameters are basically fixed: $\nu_{\max}$ is known within a factor three, $S_{\nu_{\max}}$ within 50% and the source size within a factor 10. This means that models advocating very high electron Lorentz factors ($\gamma_e \sim 10^4$, Kundt 1990) deviate from equipartition by $\sim$ 10 orders of magnitude!

Because of the high compactness of Sgr A* synchrotron self-absorption becomes another important point to be considered. Using an absorption coefficient of $\kappa_{\mathrm{sync}} = 1.4 \cdot 10^{-9} \mathrm{cm}^{-1} (n_e/\mathrm{cm}^{-3})(B/\mathrm{G})\gamma_e^{-5}(\nu/\nu_c)^{-5/3}$ one finds the synchrotron self-absorption frequency to be

$$\nu_{\mathrm{ssa}} = \frac{2.5\,\mathrm{GHz}}{k^{0.09}} \left(\frac{F_{\nu_{\max}}}{3.5\mathrm{Jy}}\right)^{0.69} \left(\frac{\nu_{\max}}{10^{12}\mathrm{Hz}}\right)^{-.46} \left(\frac{R}{10^{13}\mathrm{cm}}\right)^{-.77} \left(\frac{Z}{4\cdot 10^{13}\mathrm{cm}}\right)^{-.69}.$$

Here we took the *maximum* sizes allowed by mm-VLBI; if further studies show that Sgr A* is even more compact at submm then $\nu_{\text{ssa}}$ will increase further making it completely impossible to describe the whole spectrum with a single component.

## 3.2. SUBMM SOURCE SIZE

We can now make very solid arguments about the possible source size of Sgr A* at submm wavelengths. As VLBI measurements are only available at higher wavelengths one could still postulate arbitrarily large submm source sizes. However, if Sgr A*(submm) were optically thin and larger than $4 \cdot 10^{13}$ cm we should have seen the low frequency $\nu^{1/3}$ part of its spectrum with 3mm VLBI already. This could only be avoided if the submm component becomes optically thick below $\sim 100$ GHz. As shown above this is possible only for a very compact source where the dimensions of Sgr A* at submm wavelengths are substantially *smaller* than at $\lambda$3mm. *Consequently Sgr A* has to be equal or smaller at submm wavelengths than at $\lambda$3mm.*

Once $\nu_{\text{ssa}}$ can be determined, e.g. from broadband variability studies, we can specify the compactness of Sgr A* from its spectral characteristics alone. Arguing that the bulk of the emission at submm and mm wavelengths comes from two separate components, i.e. requiring $\nu_{\text{ssa}} \sim 100$ GHz for the submm component, would imply a source size of only

$$R \sim 1.5 \cdot 10^{12} \text{cm} \quad k^{-1/17} \left(\frac{F_{\nu_{\text{max}}}}{3.5 \text{Jy}}\right)^{8/17} \left(\frac{\nu_{\text{max}}}{10^{12} \text{Hz}}\right)^{-16/51} \left(\frac{\nu_{\text{ssa}}}{100 \text{GHz}}\right)^{-35/51}$$

for $R \sim Z$. This corresponds to $5R_{\text{g}}$ ($= 5\frac{GM_\bullet}{c^2}$) of a $2 \cdot 10^6 M_\odot$ black hole and hence to the innermost parts of an accretion disk or the very base of a jet. The fact that the non-thermal spectrum cuts-off towards the IR indicates that the submm regime indeed corresponds to the smallest spatial scale. Do we touch the supermassive black hole at these wavelengths directly?

## 3.3. MULTIPLE COMPONENTS

Although the single, monoenergetic, homogenous blob hypothesis clearly is the simplest description it appears not to be sufficient to explain Sgr A* and there are several observational indications suggesting a non-homogenous source structure, i.e.

▷ different core sizes at $\lambda$7mm and $\lambda$3mm (Krichbaum et al. 1994)
▷ different variability at radio and submm (Zylka et al. 1994)
▷ varying simultaneous spectral indices (Wright & Backer 1993).

Thus inhomogenous models (with gradients in size, $B$ and $n_{\text{e}}$, e.g. in a jet or an accretion disk) are required to describe Sgr A*.

## 4. Spherical wind accretion models

If we now want to go beyond a mere description of Sgr A*, we have to ask how this source is powered and what the underlying engine producing the radio and x-ray emission actually is? One idea is that if Sgr A* is a black hole it should swallow some fraction of the strong stellar winds seen in the GC through Bondi-Hoyle accretion.

The rate of infall depends only on the mass of the black hole and the wind parameters. Once we know the latter we can determine the black hole mass from the estimated accretion rate, which in turn could be derived from the spectrum of Sgr A*. The general validity of the Bondi-Hoyle accretion (without angular momentum) under these assumptions was recently demonstrated by 3D numerical calculations (Ruffert & Melia 1994) and the main uncertainties are related to the plasmaphysical effects associated with the infall. It is usually assumed that the magnetic field in the accreted plasma is amplified by compression up to a point where it reaches the equipartition value. Beyond this point the excess magnetic field is assumed to be dissipated and used to heat the plasma. The electron temperature is determined by the equilibrium between heating and cooling via cyclo-synchrotron radiation where one has to consider two domains for the solution of this problem: (1) hot electrons, where the typical electron Lorentz factors are of the order 100-1000 and (2) warm electrons, where the electron Lorentz factor is still close to unity.

The first domain is in a regime where synchrotron emission is important and also very effective. This requires only low accretion rates ($\dot{M} \sim 10^{-10} M_\odot/\mathrm{yr}$) and hence permits only moderately high black hole masses of the order $M_\bullet \simeq 10^3 M_\odot$ (Ozernoy 1992). The second domain is in the transition regime between cyclotron and synchrotron radiation, which is less effective than pure synchrotron radiation and hence requires higher accretion rates ($\dot{M} \sim 10^{-4} M_\odot/\mathrm{yr}$) and a higher black hole mass of the order $M_\bullet \simeq 10^6 M_\odot$ (Melia 1992 & 1994).

The big advantage of the wind-accretion approach is that it, firstly, appears unavoidable and, secondly, self-consistently ties observable parameters and accretion rate to the mass of the central object. The radio spectrum is well reproduced and initially Melia also was able to account for the x-ray flux.

On the other hand there are several counter arguments to be considered: Firstly, it is not at all clear that the wind has zero angular momentum, which would diminish the accretion rate and lead to a circularization of the accretion flow further away from the central object. There also could be residual angular momentum in Sgr A* itself, e.g. because of a fossil accretion disk which could catch the inflow further out, filling a reservoir of rather dense matter instead of directly feeding the black hole. The viscous time scales of such a disk can be very long – up to $10^7$ years (Falcke & Heinrich 1994).

There are also problems specific to each model. Ozernoy predicts a very compact source which, as shown above, would become self-absorbed already at high radio frequencies and hence requires the presence of other emission components. Melia on the other hand needs a very high accretion rate and, as Ruffert & Melia (1994) have shown, fluctuations will always lead to the formation of an accretion disk close to the black hole even for the case of initially zero angular momentum. As most of the energy of an accretion disk is produced very close to the black hole it seems impossible to avoid a high luminosity output from this accretion process. The luminosity produced by a Schwarzschild hole ($R_\mathrm{in} = 6 R_\mathrm{g}$) is $L_\mathrm{disk} = 0.8 \cdot 10^8 L_\odot \dot{M}/(10^{-4} M_\odot/\mathrm{yr})$ and even if the outer disk radius is only two times larger than $R_\mathrm{in}$, $L_\mathrm{disk}$ reduces only by a factor 3. Given the strong limits on the luminosity of Sgr A* of $L_\mathrm{disk} \ll 10^6 L_\odot$ it is very unlikely that such a high accretion rate is *currently* flowing onto the black hole. Finally, the recent SIGMA results (Goldwurm et al. 1994) are in clear contradiction with the predicted X-ray spectrum of the Melia model.

## 5. Jet-disk models

5.1. THE BASIC IDEA

Already in 1980 Reynolds and McKee argued that it is very difficult to confine the synchrotron emitting particles in Sgr A* and proposed a wind or jet model to explain the radio spectrum. Rees (1982) tried to explain Sgr A* by accretion from the interstellar matter as discussed in the previous section, however, invoking an accretion disk where the synchrotron emission stems from a relativistic electron gas in its inner parts.

We recently suggested to consider a coupled jet disk system for Sgr A* (Falcke et al. 1993a&b, Falcke & Biermann 1994a&b). The basic concept behind this approach – which has also successfully been applied to AGN – is to postulate a fundamental symbiosis between jets and disks around compact objects, i.e. that both always exist and both are energetically important. As the typical escape speed close to a black hole is scale invariant and always a large fraction of $c$ we expect at least mildly relativistic outflows irrespective of the black hole mass. The power of the jet should be mainly governed by the accretion rate.

We extended the classic Blandford & Königl (1979) jet-emission model by adding mass and energy conservation in a jet-disk system also defining scale invariant paramters for the plasma flow. A more refined model spectrum which includes the effects of adiabatic losses and non-conical jet geometry (see Reynolds 1982) but uses the same basic principles is shown in Fig. 1. Here we also accounted for the presence of a cylindrical region at the base of the jet which we termed 'nozzle', assuming that this is the region where the jet is accelerated and the electrons are injected. Hence the spectrum consists of three regions:

a) the *nozzle*, dominated by a single, quasi monoenergetic electron distribution producing the *submm* bump;
b) the *jet* itself, producing an inverted radio spectrum at *cm* wavelengths where the exact spectral index depends on the jet shape and
c) an *intermediate* region at *mm* wavelengths where both contribute equally.

The turnover frequencies between those regions depend on the self-absorption frequency of the submm component and as discussed above on the source size of jet and nozzle. Therefore one expects these parameters to be fixed by either mm-submm VLBI or simultaneous variability studies at cm-submm wavelengths.

The main finding of this kind of model is that size and flux of Sgr A* are compatible with it being a radio jet, i.e. the low accretion rate results in a very compact jet but still can yield a 1 Jy source. Although the overall power of the jet is fairly low due to the low accretion rate, the ratio between jet power $Q_{\rm jet}$ and $L_{\rm disk}$ appeared relatively high ($\sim 0.3 - 1$). This can easily be checked by crudely estimating the magnetic luminosity of Sgr A* which is $L_{\rm B} \sim 0.125(10\,{\rm G})^2 (10^{13}{\rm cm})^2 c \sim 10^4 L_\odot$. Now, one only has to remember that the total jet power including relativistic particles and kinetic energy is at least 3-4 times higher and that probably $L_{\rm disk} \le 10^5 L_\odot$.

5.2. THE AGN CONNECTION - THE CASE FOR HADRONIC CASCADES

We found that the same kind of model can not only explain Sgr A* but also the jets in AGN and even account for the tight UV-radio correlation in radio weak quasars (Falcke et al. 1994b).

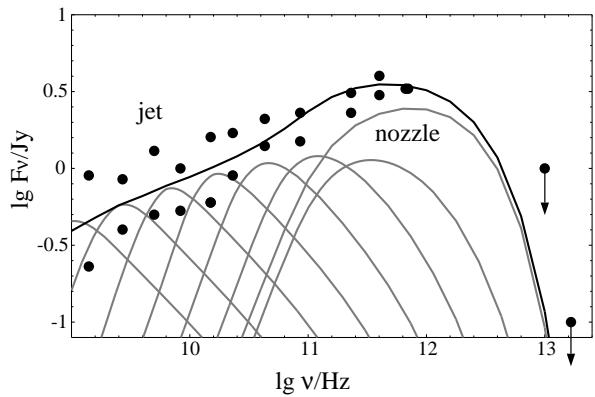

*Figure 1.* Model spectrum for jet and nozzle coupled to an accretion disk in Sgr A*. Parameters are: $R_{\rm nozz} = 3 \cdot 10^{11}$ cm, $Z_{\rm nozz} = 4.25 \cdot 10^{12}$ cm, $\gamma_{\rm e} = 70$, $q_{\rm j/l} = 0.35$, $L_{\rm disk} = 10^{39}$ erg/sec, $x_{\rm e} = 1$, $\gamma_{\rm j} = 2$, $i = 60°$ (see Falcke & Biermann 1994b). We included adiabatic losses, a nozzle where electrons are injected monoenergetically and a jet with shape $R_{\rm j} = R_{\rm nozz} + (Z_{\rm j}/Z_{\rm nozz})^{0.55}/\mathcal{M}$ slowly reaccelerating electrons into a $p = 2.5$ powerlaw.

Once more the limits imposed by the accretion disk played a crucial rôle. Again one infers injection of relativistic electrons (positrons) at high energies above $\gamma_{\rm e} = 100$ for radio loud jets and we argued that perhaps the difference between radio loud and radio weak quasars could be understood by the lack of this efficient injection mechanism in radio weak quasars (Falcke, Gopal-Krishna, Biermann 1994a). Anyway, the similarity of the high electron Lorentz factors found (directly) in Sgr A* and (indirectly) in AGN is more than striking. Hence we suggested that this typical Lorentz factor has a basic physical reason, namely the $\pi$-decay following hadronic cascades initiated by *pp*-collisions between relativistic protons in the jet and thermal protons surrounding the jet. Because of the high rest mass of the $\pi$ the secondary pairs produced in the cascade will have a characteristic energy of $> 35 MeV$ ($\gamma_{\rm e} > 70$). Jets interacting with a dense medium can inject additional high energy secondary electrons and become radio loud, while those which do not interact remain radio weak with only primary electrons injected at thermal energies – in this respect Sgr A* is *radio loud*. The latter remains true if one extends the $L_{\rm disk}$-radio correlation of AGN to lower luminosities and includes nearby Galaxies with detected radio cores and even stellar mass black holes (Falcke 1994, Falcke & Biermann 1994c): again one finds something like a radio loud/radio weak dichotomy, smoothly connecting to AGN, with Sgr A* beeing fairly loud.

Where exactly those *pp*-collisions might occur in Sgr A* is still uncertain: they may happen in an interaction zone between the jet and infalling wind or the dense absorbing material discovered by ROSAT (Predehl & Trümper 1994), but even the disk or the wind (Mastichiadis & Ozernoy 1994) itself could be a site for proton (shock-)acceleration. If *pp*-collisions are the dominant cooling process for relativistic protons being accelerated in a dense medium this would naturally yield monoenergetic secondary electrons. Below the $\pi$-production threshhold at 140 MeV *pp*-collisions are inelastic and neither produce secondaries nor lead to cooling of the protons. Once the protons are accelerated above the threshhold energy for $\pi$-production, they will instantaneously cool by *pp*-collisions until they fall below the threshhold energy thus bouncing back and forth around this energy. The resulting secondary electrons would be injected in a narrow energy interval at roughly 1/4 of the the *threshold* energy yielding $\gamma_{\rm e} \geq 70$.

## 6. Summary

Considering the dynamical and spectral evidences I have no doubt that indeed Sgr A* is the very center of the Galaxy and hence will have the coordinates $l_3 = 0$ and $b_3 = 0$ after the next revision of the galactic coordinate system (to be proposed at a future IAU assembly). Current observational data constrain models for Sgr A* already much stronger than for any other galactic nucleus – we will never get closer to a supermassive black hole. Although many question are still disputed, there is now some consensus that Sgr A* is currently put on a starvation diet – despite its high mass and strong stellar winds in the surroundings. A coupled jet/disk system can explain the spectral and structural characteristics of Sgr A* quite well and its smallest source size is close to the typical size of a black hole of mass $M_\bullet \sim 10^6$, while the typical electron Lorentz factor of $\gamma_e \sim 100$ may be indicative of hadronic cascades. Crucial future experiments will be simultaneous variability studies and mm-submm VLBI observations. Both, however, will require joined efforts to face a single but promising challenge – *understanding Sgr A*.*

## Questions

**C. Townes:** There are strong stellar winds in the GC. How can you avoid a high mass inflow – like in the Melia model – if Sgr A* is a supermassive black hole?

**Answer:** Given the bolometric luminosity constraints for UV and X-rays, I think that accretion rates as high as $10^{-4} M_\odot$/yr are already ruled out by observations. Why Sgr A* does not accrete more matter remains a mystery. Obviously we do not yet understand angular momentum distribution and transport in the inner 0.1 pc around Sgr A*. On the other hand I can not completely rule out that Sgr A* is less massive than we think it is.

**C. Townes:** Does the mass of Sgr A* play an important rôle in your models.

**Answer:** Not really. Besides the dynamical estimates, only the fact that the limits for the submm source size – giving the smallest scale – are so close to what we expect for a $10^6 M_\odot$ black hole seems very suspicious. A sign for a low mass black hole would be thermal x-ray emission from an accretion disk and heating of the ambient gas.

**T. Hasegawa:** Do you have any comments on the accretion history of the black hole? Do we see any signs of episodes of higher accretion rate in the past, or has it been starving from the very beginning of its formation?

**Answer:** There is a weak feature – the so called GC spur (Sofue, Reich & Reich 1989, ApJ 341, L47) – which could be the smoke trail of past jet activity. A single giant molecular can turn the GC into a Seyfert nucleus at any time and this could have happened already in the past. If the winds of the surrounding stars really are captured by a fossil accretion disk around Sgr A* and are stored in a close orbit than this could also lead to recurrent activity on a time scale of $10^5 - 10^7$ years.